\documentclass[prl,superscriptaddress,twocolumn,dvips]{revtex4}
\usepackage{feynmp}
\usepackage{amsmath}
\usepackage{amssymb}
\usepackage{epsfig}
\usepackage{graphicx}
\usepackage{subfigure}
\usepackage{hyperref}
\usepackage{bbm}
\usepackage{color}
\usepackage{longtable}
\usepackage{booktabs}
\usepackage{multirow}
\usepackage{array}

\newcommand{\Rmnum}[1]{\expandafter\@slowromancap\romannumeral #1@}
\allowdisplaybreaks[3]

\begin{document}

\title{Correlated interaction effects in an anisotropic flat band fermion system}

\author{Jing-Rong Wang}
\altaffiliation{Corresponding author: wangjr@hmfl.ac.cn}
\affiliation{High Magnetic Field Laboratory of Anhui Province,
Chinese Academy of Sciences, Hefei 230031, China}

\author{Chang-Jin Zhang}
\altaffiliation{Corresponding author: zhangcj@hmfl.ac.cn}
\affiliation{High Magnetic Field Laboratory of Anhui Province,
Chinese Academy of Sciences, Hefei 230031, China}
\affiliation{Institute of Physical Science and Information
Technology, Anhui University, Hefei 230601, China}

\begin{abstract}
An anisotropic flat band fermion system with a novel dispersion that is linear along one direction and cubic along another is proposed in Phys. Rev. X. 13, 021012 (2023).
We study the effects of Coulomb interaction in this fermion system by renormalization group theory and Dyson-Schwinger gap equation.
We perform renormalizaton group analysis and find that fermion velocity is always restored along the direction that the fermions take cubic dispersion originally.
Accordingly, the system takes the similar behaviors to the two-dimensional Dirac fermion system with Coulomb interaction in the low energy regime.
Based on Dyson-Schwinger gap equation method,
we find that an excitonic gap is generated if the Coulomb strength is large enough, and the system becomes a novel excitonic Chern insulator with quantized anomalous Hall conductivity.
Observable quantities of this system in free case,  under weak and strong enough Coulomb interaction are all analyzed.
\end{abstract}

\maketitle

\emph{\textbf{Introduction.}} Dirac materials attract extensive studies in the past two decades \cite{CstroNeto09, Kotov12, Vafek14, Wehling14, Hasan10, QiXL10, Armitage18, LvBQ21, Hasan21}.
Graphene is a typical two-dimensional (2D) Dirac semimetal \cite{CstroNeto09, Kotov12, Vafek14, Wehling14}. The surface state of
three-dimensional (3D) topological insulator is another typical 2D Dirac semimetal \cite{Vafek14, Wehling14, Hasan10, QiXL10}. For 2D Dirac semimetal, the Fermi surface is discrete point and density of
states takes the behavior $\rho(\omega)\propto\omega$, which vanishes at the Fermi level, i.e. $\rho(0)=0$. Accordingly, the theoretical studies showed that short-range four-fermion interactions
are irrelevant if the interactions are weak, but could drive various quantum phase transitions if the interactions are strong enough \cite{Kotov12, Herbut06, Herbut09}. Whereas, the Coulomb interaction
is long-ranged due to the vanishing of density of states. It was shown that Dirac fermion velocity in 2D Dirac semimetal increases logarithmically
with lowering of momenta under long-range Coulomb interaction \cite{Kotov12, Gonzalez94, Hofmann14, Stauber17}. The singular renormalization of fermion velocity has been observed in graphene through many different experiments
\cite{Elias11, Siegel11, Yu13}. There has been also experimental evidence
for the singular renormalization of fermion velocity for the surface state of 3D topological insulator \cite{Miao13}.

Turning some parameter properly, the fermion velocity of Dirac materials becomes to vanish and the band becomes flat
\cite{Bistritzer11, Tarnopolsky19, Cano21, WangTaiGe21, Andrei2020, Balents20, Carr20, Lau22, Mark22, Torma22, Nuckolls24, Pantaleon23}. For twisted bilayer graphene, the Dirac fermion velocity vanishes
if the twist angle between graphene layers is tuned to magic angles \cite{Bistritzer11, Tarnopolsky19}. For the surface state of
3D topological insulator, it was shown that the Dirac fermion velocity vanishes if periodic potential is applied \cite{Cano21, WangTaiGe21}.
Flat band also emerges in twisted multilayer graphene, twisted transition metal dichalcogenide, and other moire materials \cite{Andrei2020, Balents20, Carr20, Lau22, Mark22, Torma22, Nuckolls24}.
For Bernal bilayer graphene and rhombohedral multilayer graphene, the band becomes flat and the
energy dispersion of fermions takes the form $E\propto\pm k^{N}$, where $N$ is number of layers \cite{McCann13, Min08, Koshino09, ZhangFan10}.
Accordingly, the density of states satisfies $\rho(\omega)\propto\omega^{\frac{2}{N}-1}$, which
is obviously enhanced comparing to monolayer graphene. In proper conditions, higher-order van Hove point appears in the flat band fermion
system \cite{Shtyk17, Yuan19, Yuan20, Chandrasekaran20, Classen24}.

For a fermion system with flat band, the kinetic energy is suppressed, and thus correlation effects are usually much enhanced
\cite{Andrei2020, Balents20, Carr20, Lau22, Mark22, Torma22, Nuckolls24, Pantaleon23}. Many novel strong correlated phenomena, including Mott insulator \cite{CaoYuan2018B, ChenGuoRui19Mott, TangYanHo20},
unconventional superconductivity \cite{CaoYuan2018A, Yankowitz19, LuXiaoBo19, ChenGuoRui19, Arora20, Park21, HaoZeYu21, ZhouHaoXin21, ZhouHaoXin22, XiaSiYu24, GuoY24, HanTongHang24},
non-Fermi liquid behaviors \cite{Polshyn19, CaoYuan2020}, various quantum phase transitions \cite{Andrei2020, Balents20, Carr20, Lau22, Mark22, Torma22, Nuckolls24, Pantaleon23},
fractional quantum anomalous Hall effect
\cite{Park23, XuFan23, LuZhengGuang24} \emph{etc.}, are found in the flat band fermion systems.

Recently, Sheffer \emph{et al.} found that an anisotropic flat band emerges in the surface state of 3D topological insulator if  $C_{4}$ broken periodic potential is imposed properly \cite{Scheffer23}.
They showed that the Dirac fermion velocity holds along one direction and vanishes along another direction. Accordingly, the fermion dispersion is linear along one direction and cubic
along the another. They raised a very interesting question about the behaviors of this system if interactions are considered, which yet to be resolved.

In this Letter, we study the influence of Coulomb interaction on this anisotropic flat band fermion system with a linear-cubic fermion dispersion. This fermion system can be
also regarded as an anisotropic semimetal state. In this system, the density of states takes the form $\rho(\omega)\propto\omega^{\frac{1}{3}}$ which vanishes that the Fermi level. Thus, the Coulomb interaction in this system is long-ranged, which could  induce strong-correlated phenomena.

Influence of long-range Coulomb interaction on various semimetal state attracted broad interest \cite{Goswami11, Hosur12, Moon13, Herbut14, YangNatPhys14, Abrikosov72, Isobe16, Cho16, WangLiuZhang17A, Lai15, Jian15, WangLiuZhang17B, ZhangShiXin17, WangLiuZhang18, WangLiuZhang19, Han19, ZhangSX21, Roy18Birefringent, Kotov20, LeeYuWen21, Roy23}. These studies manifest that the results subtly depend on the dispersion of
fermions and dimension of the system, and thus need careful analysis. In the following, we study the influence of long-range Coulomb interaction on this anisotropic flat band fermion system by renormalization group (RG)
theory \cite{Shankar94} and self-consistent Dyson-Schwinger gap equation. We reveal the behaviors of the system not only in presence of weak Coulomb interaction but also  with  strong Coulomb
interaction.

\begin{figure}[htbp]
\center
\includegraphics[width=3.3in]{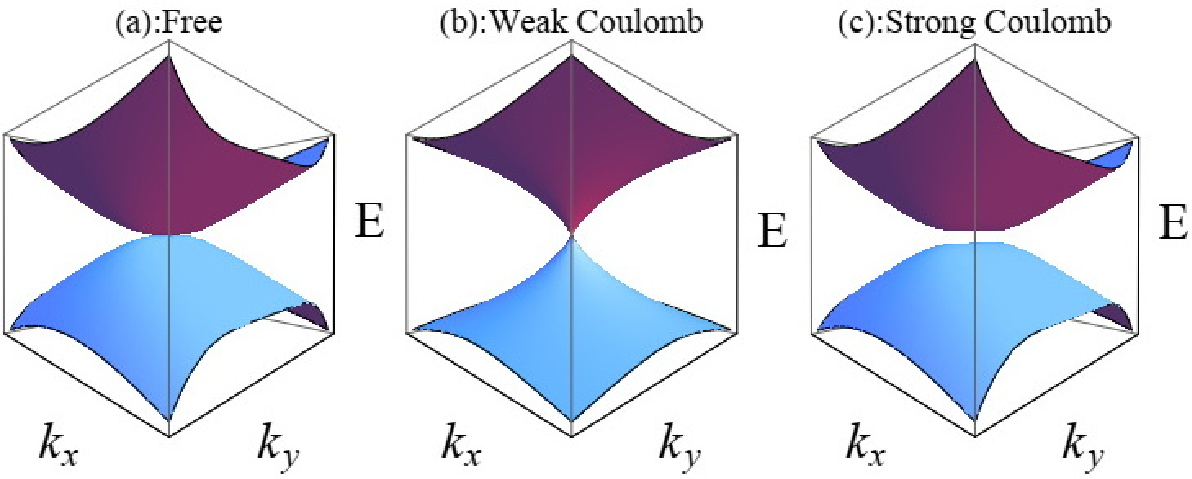}
\caption{Fermion dispersion of anisotropic band with  linear-cubic fermion (a) in free case, (b) with weak Coulomb interaction becomes, and (c)
 with strong enough Coulomb interaction.     \label{Fig:Dispersion}}
\end{figure}

\emph{\textbf{The model.}} The action of free fermions can be written as
\begin{eqnarray}
S_{\psi} &=& \int\frac{d\omega}{2\pi}\frac{d^2\mathbf{k}}{(2\pi)^{2}}
\psi^{\dag}\left(-i\omega+\mathcal{H}_{f}(\mathbf{k})\right)\psi ,
\end{eqnarray}
where the Hamiltonian for the the fermions is
\begin{eqnarray}
\mathcal{H}_{f}(\mathbf{k})=v_{x}k_{x}\sigma_{1}+d_{y}k_{y}^{3}\sigma_{2}.
\end{eqnarray}
The field operator $\psi$ is a
two-component spinor. $\sigma_{i}$ with $i=1,2,3$ are the Pauli matrices. $v_{x}$ and $d_{y}$ are the
model parameters. The fermion energy dispersion is given by
\begin{eqnarray}
E=\pm\sqrt{v_{x}^2k_{x}^{2}+d_{y}^{2}k_{y}^{6}},
\end{eqnarray}
which is linear along one direction, and cubic along another direction.

The Coulomb interaction between fermions takes the form
\begin{eqnarray}
H_{C} = \frac{1}{4\pi}\int
d^2\mathbf{x}d^2\mathbf{x}'\rho(\mathbf{x})\frac{e^{2}}{\epsilon
\left|\mathbf{x}-\mathbf{x}'\right|}\rho(\mathbf{x}'),
\end{eqnarray}
where $\rho(\mathbf{x})=\psi^{\dag}(\mathbf{x})\psi(\mathbf{x})$
is fermion density operator, $e$ electric charge, and $\epsilon$
dielectric constant.

The free fermion propagator reads
\begin{eqnarray}
G_{0}(\omega,\mathbf{k}) = \frac{1}{-i\omega+v_{x}k_{x}\sigma_{1}+d_{y}k_{y}^{3}\sigma_{2}}.\label{Eq:FreeFermonPropagatorDef}
\end{eqnarray}
The bare Coulomb interaction is written in the momentum space as
\begin{eqnarray}
V_{0}(\mathbf{q}) = \frac{2\pi e^2}{\epsilon|\mathbf{q}|} =
\frac{2\pi\alpha v}{|\mathbf{q}|},
\end{eqnarray}
where $\alpha = e^2/\epsilon v_{x}$ represents the effective interaction
strength.

The polarization is defined as
\begin{eqnarray}
\Pi(\Omega,\mathbf{q})
&=&-\int\frac{d\omega}{2\pi}\int\frac{d^2\mathbf{k}}{(2\pi)^{2}}
\mathrm{Tr}\left[G_{0}(\omega,\mathbf{k})\right.\nonumber
\\
&&\left.\times G_{0}\left(\omega+\Omega,\mathbf{k}+\mathbf{q}\right)\right].
\end{eqnarray}
Substituting the fermion propagator, we find the polarization can be approximately by \cite{Supplement}
\begin{eqnarray}
\Pi(\Omega,q_{x},q_{y})
&=&\frac{1}{v_{x}d_{y}^{\frac{1}{3}}}\left[\frac{c_{1}v_{x}^{2}q_{x}^{2}}{\left(\Omega^{2}+v_{x}^{2}q_{x}^{2}+\frac{c_{3}^{6}}{c_{2}^{6}}d_{y}^{2}q_{y}^{6}\right)^{\frac{5}{6}}}\right.\nonumber
\\
&&\left.+\frac{c_{3}d_{y}^{\frac{2}{3}}q_{y}^{2}}{\left(\Omega^{2}+v_{x}^{2}q_{x}^{2}+\frac{c_{3}^{6}}{c_{2}^{6}}d_{y}^{2}q_{y}^{6}\right)^{\frac{1}{6}}}\right],\label{Eq:ExpPolarization}
\end{eqnarray}
where
\begin{eqnarray}
c_{1}&=&\frac{\Gamma\left(\frac{1}{6}\right)}{24\cdot 2^{\frac{1}{3}}\sqrt{\pi} \Gamma\left(\frac{2}{3}\right)}, \qquad
c_{2}=\frac{17}{144}, \nonumber
\\
c_{3}&=&\frac{9\Gamma\left(\frac{5}{6}\right)}{32\cdot 2^{\frac{2}{3}}\sqrt{\pi}\Gamma\left(\frac{4}{3}\right)}.
\end{eqnarray}

\begin{figure*}[htbp]
\center
\includegraphics[width=6.6in]{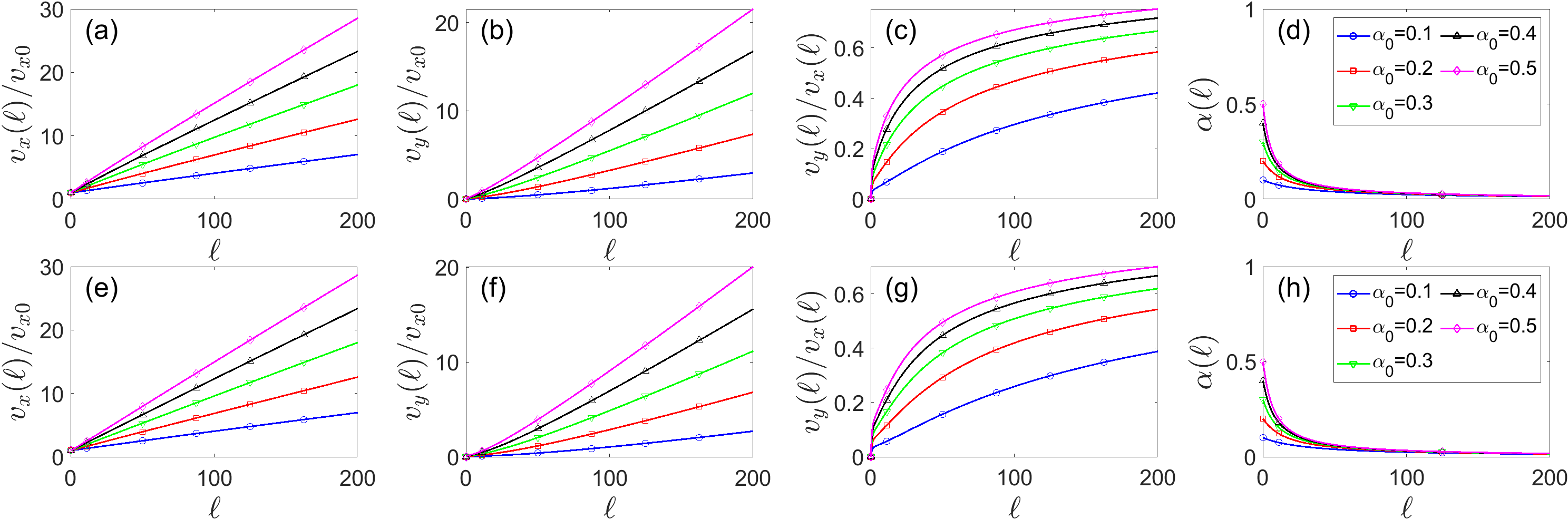}
\caption{ Flows of $v_{x}(\ell)$, $v_{y}(\ell)$, $v_{y}(\ell)/v_{x}(\ell)$ and $\alpha(\ell)$ with different initial values of Coulomb strength. (a)-(d) Bare propagator of Coulomb interaction is employed.
(e)-(f) Dressed Coulomb interaction is used. Initial value $\beta_{0}=1$ is taken. \label{Fig:VRG}}
\end{figure*}

\emph{\textbf{Restoring of fermion velocity.}} In the following, we analyze the effects of Coulomb interaction through RG method \cite{Shankar94}.
Firstly, we calculate the fermion self-energy neglecting the screen from polarization function.
The fermion self-energy can be written as
\begin{eqnarray}
\Sigma(\omega,\mathbf{k})=\int\frac{d\Omega}{2\pi}
\int'\frac{d^2\mathbf{q}}{(2\pi)^{2}}
G_{0}\left(\omega+\Omega,\mathbf{k}+\mathbf{q}\right)V_{0}(\Omega,\mathbf{q}),
\end{eqnarray}
where $\int'$ represents that a momentum shell will be imposed properly.
Substituting the propagator of fermion as shown in Eq.~(\ref{Eq:FreeFermonPropagatorDef}), and
expanding to leading terms of $k_{x}$ and $k_{y}$, we obtain
\begin{eqnarray}
\Sigma(\omega,\mathbf{k})&=&-i\omega\Sigma_{0}+v_{x}k_{x}\sigma_{1}\Sigma_{1}+d_{y}k_{y}^{3}\sigma_{2}\Sigma_{2}\nonumber
\\
&&+k_{y}\sigma_{2}\Sigma_{L},
\end{eqnarray}
where the expressions of $\Sigma_{0}$, $\Sigma_{1}$, $\Sigma_{2}$, $\Sigma_{L}$  can be found in the Supplementary Materials \cite{Supplement}. We notice that linear term of
$k_{y}$ is always generated dynamically. According, the fermion velocity along $y$ axis is restored.

In order to consider this dynamically generated term in RG analysis, we employ the fermion propagator as following
\begin{eqnarray}
G_{0}^{new}(\omega,\mathbf{k}) = \frac{1}{-i\omega+v_{x}k_{x}\sigma_{1}+v_{y}k_{y}\sigma_{2}+d_{y}k_{y}^{3}\sigma_{2}}.\label{Eq:FreeFermonPropagatorNew}
\end{eqnarray}
The bare value of $v_{y}$ is taken to be zero in the RG analysis.
The fermion self-energy can be written as
\begin{eqnarray}
\Sigma(\omega,\mathbf{}k)&=&\int\frac{d\Omega}{2\pi}
\int'\frac{d^2\mathbf{q}}{(2\pi)^{2}}
G_{0}^{new}\left(\omega+\Omega,\mathbf{k}+\mathbf{q}\right)\nonumber
\\
&&\times V_{0}(\Omega,\mathbf{q}).
\end{eqnarray}
Substituting the propagator of fermion as shown in Eq.~(\ref{Eq:FreeFermonPropagatorNew}), and expanding to the leading terms of $\omega$, $k_{x}$ and $k_{y}$, we arrive 
\begin{eqnarray}
\Sigma(\omega,\mathbf{k})
&=&v_{x}k_{x}\sigma_{1}C_{1}\ell+v_{y}k_{y}\sigma_{2}C_{2}^{a}\ell+d_{y}k_{y}^{3}\sigma_{2}C_{2}^{b}\ell.
\end{eqnarray}
$C_{1}$, $C_{2}^{a}$, and $C_{2}^{b}$ are functions of the parameters $\alpha$, $\beta$, and $\gamma$, which are defined as
$\alpha=\frac{e^{2}}{\epsilon v_{x}}$, $\beta=\frac{d_{y}\Lambda^{2}}{v_{x}}$, $\delta=v_{y}/v_{x}$.
The expressions of $C_{1}$, $C_{2}^{a}$, and $C_{2}^{b}$ are shown in \cite{Supplement}.
The RG scheme $-\infty<\Omega<+\infty$, $b\Lambda<|\mathbf{q}|<\Lambda$ has been utilized,
where $b=e^{-\ell}$ with $\ell$ being the RG running parameter. We get the RG equation as following
\begin{eqnarray}
\frac{dv_{x}}{d\ell}&=&C_{1}v_{x}, \label{Eq:VRGvx}
\\
\frac{dv_{y}}{d\ell}&=&C_{2}^{a}v_{y}, \label{Eq:VRGvy}
\\
\frac{d\alpha}{d\ell}&=&-C_{1}\alpha. \label{Eq:VRGalpha}
\end{eqnarray}

The RG flows of $v_{x}(\ell)$, $v_{y}(\ell)$, $v_{y}(\ell)/v_{x}(\ell)$, and $\alpha(\ell)$ are shown in Figs.~\ref{Fig:VRG}(a)-\ref{Fig:VRG}(d).
$v_{x}(\ell)$ increases nearly linearly with increasing of $\ell$. $v_{y}(\ell)$ grows from zero and also increases nearly linearly with increasing of $\ell$. These results
indicate that the fermion velocities $v_{x}$ and $v_{y}$ acquire logarithmic-like dependence on momenta. The ratio of fermion velocities $v_{y}(\ell)/v_{x}(\ell)$ approaches
to 1 slowly with increasing of $\ell$. The result that $\alpha(\ell)$ flows to zero slowly in the low energy limit $\ell\rightarrow\infty$ represents that Coulomb interaction
is marginally irrelevant in the low energy regime.

Subsequently, we calculate the self-energy of fermions considering screen of polarization. In order to consider
the original fermion dispersion and the influence of restoring of  fermion velocity $v_{y}$. We employ the dressed Coulomb
interaction as following
\begin{eqnarray}
V^{\star}(\Omega,\mathbf{q})=\frac{1}{V_{0}^{-1}(|\mathbf{q}|)+F_{1}\Pi(\Omega,\mathbf{q})+F_{2}\Pi_{\mathrm{Dirac}}(\Omega,\mathbf{q})},
\end{eqnarray}
where $\Pi(\Omega,\mathbf{q})$ is given by (\ref{Eq:ExpPolarization}), and
\begin{eqnarray}
\Pi_{\mathrm{Dirac}}(\Omega,\mathbf{q})&=&\frac{1}{16v_{x}v_{y}}\frac{v_{x}^{2}q_{x}^{2}+v_{y}^{2}q_{y}^{2}}{\sqrt{\Omega^{2}+v_{x}^{2}q_{x}^{2}+v_{y}^{2}q_{y}^{2}}},
\\
F_{1}&=&e^{-\frac{v_{y}\Lambda}{d_{y}\Lambda^{3}}}=e^{-\frac{v_{y}}{d_{y}\Lambda^{2}}},
\\
F_{2}&=&e^{-\frac{d_{y}\Lambda^{3}}{v_{y}\Lambda}}=e^{-\frac{d_{y}\Lambda^{2}}{v_{y}}},
\end{eqnarray}
where  $\Pi_{\mathrm{Dirac}}$ is the polarization induced by 2D Dirac fermions \cite{Kotov12}.
Considering the dressed Coulomb interaction, the RG equations for $v_{x}$ and $v_{y}$ become
\begin{eqnarray}
\frac{dv_{x}}{d\ell}&=&\left(C_{1}^{\star}-C_{0}^{\star}\right)v_{x}, \label{Eq:VRGvxScreen}
\\
\frac{dv_{y}}{d\ell}&=&\left({C_{2}^{a}}^{\star}-C_{0}^{\star}\right)v_{y}, \label{Eq:VRGvxScreen}
\\
\frac{d\alpha}{d\ell}&=&\left(-C_{1}^{\star}+C_{0}^{\star}\right)\alpha,  \label{Eq:VRGvxScreen}
\end{eqnarray}
where $C_{0}^{\star}$, $C_{1}^{\star}$, $C_{2}^{\star}$, and ${C_{2}^{a}}^{\star}$ are the functions of $\alpha$, $\beta$, and $\delta$ \cite{Supplement}.
The RG flows of $v_{x}(\ell)$, $v_{y}(\ell)$, $v_{y}(\ell)/v_{x}(\ell)$ and $\alpha(\ell)$ considering of screen from polarization functions are
shown in Figs.~\ref{Fig:VRG}(e)-\ref{Fig:VRG}(h). We can find that these RG flows have the qualitatively same behaviors as the ones shown in
Figs.~\ref{Fig:VRG}(a)-\ref{Fig:VRG}(d).

\begin{figure}[htbp]
\center
\includegraphics[width=3.3in]{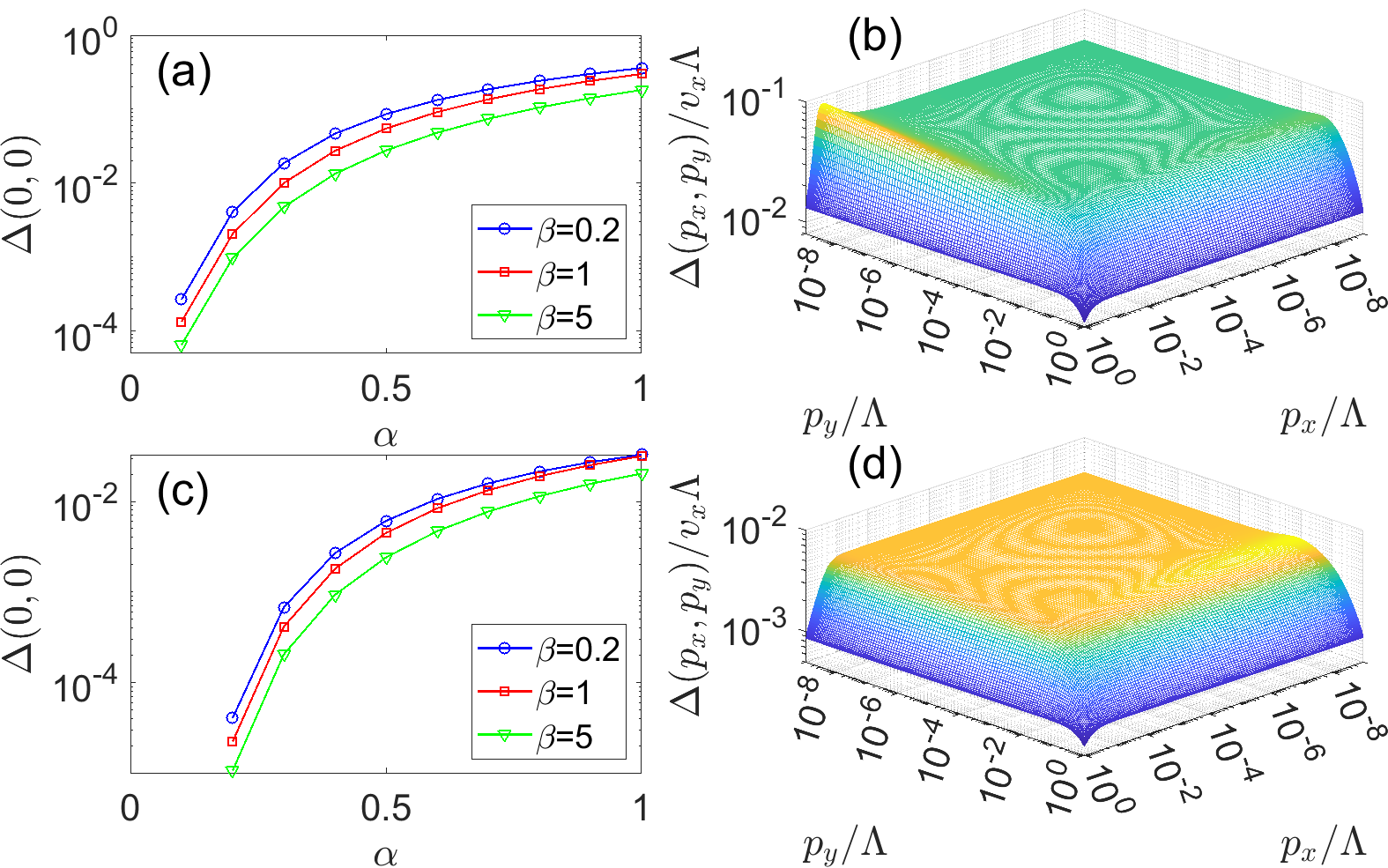}
\caption{(a) and (b) $\Delta(0,0)$ and $\Delta(p_{x},p_{y})$ calculated by bare propagator of Coulomb interaction. (c) and (d) $\Delta(0,0)$ and $\Delta(p_{x}, p_{y})$ calculated by dressed Coulomb interaction including
the screen of polarization. In (b) and (d), $\alpha=0.5$ and $\beta=1$ are taken. \label{Fig:ExcitonicGap}}
\end{figure}

\emph{\textbf{Generation of excitonic gap.}} If the Coulomb interaction is strong enough, an excitonic gap can be generated
\cite{Khveshchenko01, Gorbar02, Liu09, Gamayun10, WangLiu12, Gonzalez15, Carrington16, Carrington18, CastroNeto09, Drut09}. In order to study the generation of excitonic gap, we employ
the Dyson-Schwinger equation method \cite{Khveshchenko01, Gorbar02, Liu09, Gamayun10, WangLiu12, Gonzalez15, Carrington16, Carrington18}.
The gap equation  to lowest order can be written as
\begin{eqnarray}
\Delta(p_{x},p_{y})
&=&\frac{1}{2}\int\frac{d^2\mathbf{k}}{(2\pi)^2}\frac{\Delta(k_{x},k_{y})}
{\sqrt{v_{x}^2k_{x}^{2}
+d_{y}^2k_{y}^6+\Delta^2(\omega,k_{x},k_{y})}}\nonumber
\\
&&\times V_{0}(\mathbf{p}-\mathbf{k}). \label{Eq:GapEquationLowestOrder}
\end{eqnarray}
Through numerical calculation, we can find that excitonic gap $\Delta(p_{x},p_{y})$ is dynamically generated if the Coulomb interaction $\alpha$ is
larger enough.
Relations between $\Delta(0,0)$ and $\alpha$ with different values of $\beta$ are presented in Fig.~\ref{Fig:ExcitonicGap}(a). It is easy to notice that $\Delta(0,0)$ becomes finite if $\alpha$ is larger than a critical value.
 Dependence of $\Delta(p_{x},p_{y})$ on $p_{x}$ and
$p_{y}$ is shown in Fig.~\ref{Fig:ExcitonicGap}(b).  We find that $\Delta(p_{x},p_{y})$ approaches to a constant value $\Delta(0,0)$ when $p_{x}$ and $p_{y}$ approach to zero. $\Delta(p_{x},p_{y})$ drops
quickly when $vp_{x}$ or $d_{y}p_{y}^{3}$ is larger than the energy scale given by $\Delta(0,0)$.

Considering the correction from polarization to lowest order, the dressed Coulomb interaction becomes
\begin{eqnarray}
V(\Omega,\mathbf{q})=\frac{1}{V_{0}^{-1}(\mathbf{q})+\Pi(\Omega,\mathbf{q})},
\end{eqnarray}
where $\Pi(\Omega,\mathbf{q})$ is given by Eq.~(\ref{Eq:ExpPolarization}).
Employing the instantaneous approximation \cite{Khveshchenko01, Gorbar02, Liu09}, the Coulomb interaction becomes
\begin{eqnarray}
V(\mathbf{q})=\frac{1}{V_{0}^{-1}(\mathbf{q})+\Pi(0,\mathbf{q})}. \label{Eq:DressedCoulombInstantanesous}
\end{eqnarray}
Replacing $V_{0}(\mathbf{q})$ in Eq.~(\ref{Eq:GapEquationLowestOrder}) with $V(\mathbf{q})$ as shown in Eq.~(\ref{Eq:DressedCoulombInstantanesous}),
we obtain the gap equation considering the screening from polarization function under instantaneous approximation. Dependence of $\Delta(0,0)$ on $\alpha$ and dependence of
$\Delta(p_{x},p_{y})$ on momenta $p_{x}$ and $p_{y}$ considering screen from polarization function are presented in Figs.~\ref{Fig:ExcitonicGap}(c) and \ref{Fig:ExcitonicGap}(d)
respectively. Comparing Fig.~\ref{Fig:ExcitonicGap}(c) with \ref{Fig:ExcitonicGap}(a), and \ref{Fig:ExcitonicGap}(d) with \ref{Fig:ExcitonicGap}(b), we can find that
the magnitude of excitonic gap is suppressed if the screening from polarization is considered.

\emph{\textbf{Observable quantities.}} For free linear-cubic fermion system, the observable quantities including density of states $\rho(\omega)$, specific heat $C_{v}(T)$,
compressibility $\kappa(T)$,   diamagnetic susceptibility $\chi_{D}(T)$, and optical conductivities along the two directions $\sigma_{xx}(\omega)$ and $\sigma_{yy}(\omega)$
take the behaviors \cite{Supplement}
\begin{eqnarray}
&&\rho(\omega)\propto\omega^{1/3}, \qquad C_{v}(T)\propto T^{4/3},\nonumber
\\
&&\kappa(T)\propto T^{1/3}, \qquad  \chi_{D}(T)\propto T^{-1/3}, \nonumber
\\
&&\sigma_{xx}(\omega)\propto\omega^{-2/3},\qquad \sigma_{yy}(\omega)\propto\omega^{2/3},
\end{eqnarray}
where $\omega$ represents the energy, and $T$ is the temperature.

Under weak Coulomb interaction, the system takes the similar behaviors as 2D Dirac semimetal with Coulomb interaction. Thus, the observable quantities can be written as \cite{Kotov12, Sheehy07}
\begin{eqnarray}
\rho(\omega)\propto\frac{\omega}{\ln^{2}\left(\frac{\omega_{0}}{\omega}\right)},\qquad C_{v}(T)\propto\frac{T^{2}}{\ln^{2}\left(\frac{T_{0}}{T}\right)}, \nonumber
\\
\kappa(T)\propto \frac{T^{2}}{\ln^{2}\left(\frac{T_{0}}{T}\right)}, \qquad \chi_{D}(T)\propto \frac{\ln^{2}\left(\frac{T_{0}}{T}\right)}{T},\nonumber
\\
\sigma_{xx,yy}(\omega)\propto \sigma_{0}\left[1+\frac{\tilde{C\alpha}}{1+\alpha\ln\left(\frac{\omega_{0}}{\omega}\right)}\right].
\end{eqnarray}

If the Coulomb interaction is strong enough, an excitonic gap $\Delta$ is generated. Then the observable quantities take the behaviors \cite{Supplement}
\begin{eqnarray}
\rho(\omega)
&\propto&
\frac{|\omega|}{\left(\omega^{2}-\Delta^{2}\right)^{\frac{1}{3}}}\theta\left(|\omega|-\Delta\right), \nonumber
\\
C_{v}(T)
&\propto&\frac{\Delta^{\frac{10}{3}}}{T^{2}}
e^{-\frac{\Delta}{T}}, \nonumber
\\
\kappa(T)
&\propto&\frac{\Delta^{\frac{4}{3}}}{T}
e^{-\frac{\Delta}{T}}, \nonumber
\\
\chi(T)&\propto&\frac{1}{\Delta^{\frac{1}{3}}}, \nonumber
\\
\sigma_{xx}(\omega)
&\propto&\frac{1+12\frac{\Delta^{2}}{\omega^{2}}}{\left(\omega^2-4\Delta^2\right)^{\frac{1}{3}}}
\theta\left(|\omega|-2\Delta\right), \nonumber
\\
\sigma_{yy}(\omega)
&\propto&
\left(\omega^2-4\Delta^2\right)^{\frac{1}{3}}
\left(1+\frac{20}{3}\frac{\Delta^{2}}{\omega^{2}}\right)
\theta\left(|\omega|-2\Delta\right).
\end{eqnarray}
The behaviors of $C_{v}(T)$, $\kappa(T)$, and $\chi_{D}(T)$ are valid in the limit $T\ll\Delta$.

Considering the excitonic gap $\Delta$, the Hamiltonian can be written as
\begin{eqnarray}
\mathcal{H}_{\Delta}&=&D_{x}\sigma_{x}+D_{y}\sigma_{y}+D_{z}\sigma_{z}, 
\end{eqnarray}
where the vector $\mathbf{D}$ is defined as
\begin{eqnarray}
\mathbf{D}
&=&v_{x}k_{x}\mathbf{e}_{x}+d_{y}k_{y}^{3}\mathbf{e}_{y}+\Delta\mathbf{e}_{z}.
\end{eqnarray}
The topological property of this Hamiltonian is determined by the  unit vector
\begin{eqnarray}
\hat{\mathbf{D}}=\frac{\mathbf{D}}{|\mathbf{D}|}
=\frac{v_{x}k_{x}\mathbf{e}_{x}+d_{y}k_{y}^{3}\mathbf{e}_{y}+\Delta\mathbf{e}_{z}}
{\sqrt{v_{x}^{2}k_{x}^{2}+d_{y}^{2}k_{y}^{6}+\Delta^{2}}}.
\end{eqnarray}
The corresponding Chern number is given by \cite{QiXiaoLiang06, LiuChaoXing16}
\begin{eqnarray}
C=\frac{1}{4\pi}\int d^2\mathbf{k}\left(\frac{\partial \hat{\mathbf{D}}}{\partial k_{x}}\times \frac{\partial \hat{\mathbf{D}}}{\partial k_{y}}\right)\cdot\hat{\mathbf{D}}.
\end{eqnarray}
Substituting the expression of $\hat{\mathbf{D}}$, we can get
\begin{eqnarray}
C
&=&\frac{1}{4\pi}\int d^2\mathbf{k}
\frac{3v_{x}d_{y}k_{y}^{2}\Delta}{\left[v_{x}^{2}k_{x}^{2}+d_{y}^{2}k_{y}^{6}+\Delta^{2}\right]^{\frac{3}{2}}}\nonumber.
\\
&=&\frac{1}{2}\mathrm{sgn}(v_{x})\mathrm{sgn}(d_{y})\mathrm{sgn}(\Delta).
\end{eqnarray}
Thus, the system has a quantum anomalous Hall conductivity \cite{QiXiaoLiang06, LiuChaoXing16}
\begin{eqnarray}
\sigma_{xy}=C\frac{e^{2}}{h},
\end{eqnarray}
where $h$ is the  Planck constant.
The system becomes a novel excitonic Chern insulator under strong Coulomb interaction.

\emph{\textbf{Discussion.}} In summary, we study the influence of long-range Coulomb interaction on an anisotropic flat band fermion
with a dispersion which is linear along one direction and cubic along another. Based on RG theory,
we show that the fermion velocity is always restored along the direction which has cubic dispersion originally. Accordingly,
the system takes the qualitatively same behaviors as the 2D Dirac fermion system with Coulomb interaction. Through self-consistent
Dyson-Schwinger gap equation method, we find that the system is driven to excitonic Chern insulator phase if the Coulomb
interaction is strong enough.

In Ref.~\cite{Scheffer23}, there is a term $k_{x}^{2}k_{y}\sigma_{2}$ for the Hamiltonian. In this Letter, we neglect this term
to simplify the calculation. Our conclusions will be not changed qualitatively if this term is considered.

Due to vanishing of density of sates at the Fermi level, weak four-four interaction is irrelevant in the view of RG theory,
but strong four-fermion interaction could become relevant and drive the system to a new phase. Considering four-fermion interaction
$g_{3}\left(\psi^{\dag}\sigma_{3}\psi\right)^{2}$,  through analysis of mean-field method, we find that an excitonic gap $\Delta$ is generated if $g_{3}$ is larger than a critical
value $g_{3c}$, and it takes the form $\Delta\propto\left(g_{3}-g_{3c}\right)^{3}$ \cite{Supplement}.

Studies about the interplay of Coulomb interaction and short-range four-fermion interaction, and the quantum critical behaviors between
the gapless anistropic band fermion phase and excitonic Chern insulator phase are also interesting.

\section*{ACKNOWLEDGEMENTS}

We acknowledge the support from National Key R\&D Program of China (Grant Nos. 2022YFA1403203, 2024YFA1611103, 2021YFA1600201),
the National Natural Science Foundation of China under Grant 12274414, and the Basic Research Program of
the Chinese Academy of Sciences Based on Major Scientific
Infrastructures (Contract No. JZHKYPT-2021-08).

\end{document}